\documentclass[prl,10pt,twocolumn,showpacs]{revtex4}

\usepackage[dvips]{epsfig}
\usepackage{float}
\usepackage{amsfonts}
\usepackage{amsmath}
\usepackage{amssymb}
\usepackage{enumerate}

\begin{document}

\title{Optical fiber coupling to planar photonic crystal microcavities}

\author{Paul E. Barclay}
\email{pbarclay@caltech.edu}
\author{Kartik Srinivasan}
\author{Oskar Painter}
\affiliation{Department of Applied Physics, California Institute of Technology, Pasadena, California 91125}
\date{\today}
\begin{abstract}
A technique is demonstrated which efficiently transfers light between a tapered standard single-mode optical fiber and a resonant mode of a high-Q photonic crystal cavity with mode volume less than a cubic wavelength in size.   Cavity mode quality factors of $4.7\times 10^4$ are measured, and a total fiber-to-cavity coupling efficiency of 44\% is demonstrated.
\end{abstract}

\maketitle

\setcounter{page}{1}

Recently it has been demonstrated that resonant microcavities formed in planar photonic crystals (PC) are capable of not only confining light to ultra-small optical mode volumes\cite{ref:Srinivasan4}, but can also be made of high enough quality to enable photon cavity lifetimes capable of, for instance, reaching strong-coupling with atomic Cs\cite{ref:Lev} or semiconductor quantum dots\cite{ref:Srinivasan3,ref:Srinivasan4,ref:Noda4}.  In many of the proposed applications of such PC cavities in quantum, nonlinear, and integrated optics, the ability to efficiently interface the PC cavities with external optics is also of critical importance. In particular, using high-Q PC cavities for chip based cavity-QED (cQED) \cite{ref:Lev} or in single-photon sources \cite{ref:Gerard3}, where photon collection is an important measure of device performance \cite{ref:Brassard1}, requires an efficient coupling scheme to the sub-micron cavity mode.  The difficulty in optically accessing PC cavities is largely a result of their ultra-small mode volumes and external radiation pattern, which unlike micropost \cite{ref:Pelton} and Fabry-Perot \cite{ref:McKeever1} cavities, is not inherently suited to coupling with conventional free-space or fiber optics.  In this Letter we present an optical fiber-based coupling technique, and employ it to efficiently source and collect light from the resonant modes of a high-Q PC cavity.

\begin{figure}[ht]
\begin{center}
\epsfig{figure=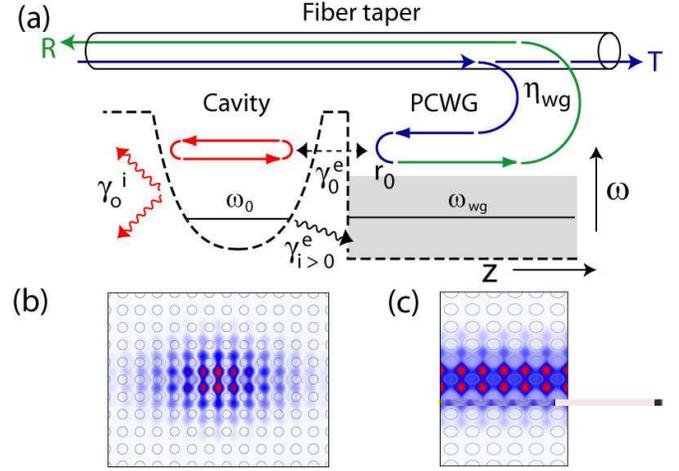, width=1.0\linewidth}
\caption{(a) Illustration of the fiber-PC cavity coupling process.  The dashed line is meant to represent the band-edge frequency  in the planar PC  as a function of the position $z$ along the waveguide axis, with the shaded region representing the bandwidth of the fundamental PCWG mode.   Magnetic field profile of (b) the high-Q PC cavity $A_{2}$ mode, and (c) the fundamental $TE_{1}$ PCWG mode.}
\label{fig:illustration}
\end{center}
\end{figure} 

An illustration of the coupling scheme is shown in Fig.\ \ref{fig:illustration}(a).  Evanescent coupling between an optical fiber taper \cite{ref:Knight} and a planar photonic crystal waveguide (PCWG) is initially used to interface with the PC chip.  Once on the chip, light is guided to a PC cavity at the terminus of the PCWG.  This PC cavity-waveguide system was previously studied in Refs. \cite{ref:Barclay2}, where the fundamental ($TE_{1}$) mode of the PCWG (Fig.\ \ref{fig:illustration}(c)) was designed to mode-match with the fundamental ($A_{2}$) cavity mode (Fig. \ \ref{fig:illustration}(b)).  The mode-matched cavity acts as a mirror with high reflectivity, $r_o(\omega)$, except at the frequencies of the localized  cavity states, where light can resonantly tunnel between the PCWG and the cavity.  The reflected signal from the PC cavity is finally recollected into the backward propagating fiber taper mode.  Previous measurements of the optical fiber-PCWG evanescent coupler have shown that near unity coupling efficiency ($97\%$) over a bandwidth of roughly 15 nm\cite{ref:Barclay5} is possible, and in an independent study, the $A_{2}$ mode of the graded square lattice PC cavity was measured to have a $Q$-factor of $4 \times 10^4$ and a mode localization consistent with an effective mode volume of $V_{\text{eff}} = 0.9(\lambda/n)^3$\cite{ref:Srinivasan4}.  In this work we aim to study the efficiency with which the PCWG can load the PC cavity using this fiber-coupled approach.

\begin{figure}[ht]
\begin{center}
\epsfig{figure=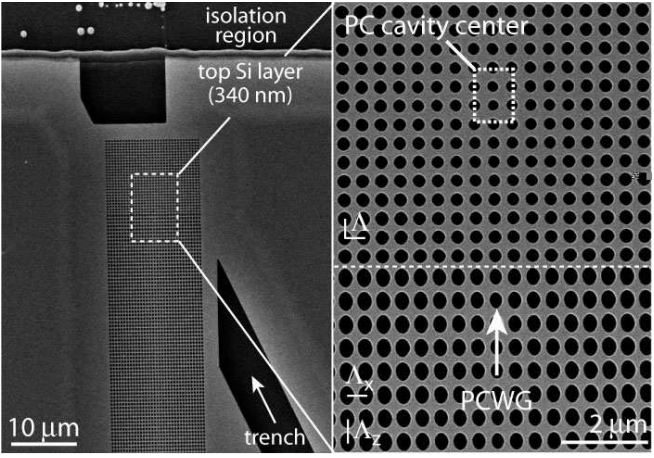, width=1.0\linewidth}
\caption{SEM image of an integrated PCWG-PC cavity sample.  The PC cavity and PCWG have lattice constants $\Lambda \sim 430$ $\text{nm}$, $\Lambda_{x} \sim 430$ $\text{nm}$, and $\Lambda_{z} \sim 550$ $\text{nm}$.  The surrounding silicon material has been removed to form a diagonal trench and isolated mesa structure to enable fiber taper probing.}
\label{fig:SEM}
\end{center}
\end{figure} 

As proposed in Ref. \cite{ref:Spillane2}, the interaction between the PC cavity and the external PCWG can be described by two key parameters, the \emph{coupling parameter} $K$ and the \emph{ideality factor} $I$:
\begin{align}
K &\equiv \frac{\gamma^{e}_{0}}{\gamma^{i}_{o} + \sum_{i \ne 0} \gamma^{e}_{i}}, \label{eq:coupling_factor}\\
I &\equiv \frac{\gamma^{e}_{0}}{\sum_{i} \gamma^{e}_{i}},\label{eq:ideality_factor}
\end{align}

\noindent where the cavity mode is characterized by its resonance frequency $\omega_o$, its intrinsic photon loss rate ($\gamma^{i}_{o}$) in absence of the external PCWG, and its coupling rates to the fundamental ($TE_{1}$ mode) and higher order (including radiating) modes of the external PCWG, $\gamma^{e}_{0}$ and $\gamma^{e}_{i>0}$, respectively.  $I$ and $K$ describe the degree of ``good'' loading (via the PCWG $TE_1$ mode in this case) relative to the  total loading of the resonator, and the parasitic and intrinsic loss channels of the resonator, respectively.  On resonance, the fraction of the optical power reflected by the cavity back into the PCWG mode is given by, $R_{o}(\omega_{o}) = (1-K)^2/(1+K)^2$.  The remaining (fractional) optical power, $1-R_{o}(\omega_{o})$, is absorbed inside the PC cavity or radiated into the parasitic output channels.  The measured reflection resonance linewidth is given by the sum of the loss rates for \emph{all} of the loss channels, $\delta\omega = \gamma^i_{o} + \sum_{i} \gamma^{e}_{i}$.  The quality factor of the PC cavity mode due to intrinsic and parasitic loss (i.e., those loss channels other than the ``good'' PCWG $TE_{1}$ channel) can then be determined from $R_{o}(\omega_{o})$ and $\delta\omega$,
\begin{equation}\label{eq:Q_i+P}
Q_{i+P} = 2Q_{T}\frac{1}{1 \pm \sqrt{R_{o}(\omega_{o})}} = Q_{T}(1+K),
\end{equation}
where the total loaded quality factor is $Q_{T} = \omega_{o}/\delta\omega$, and where the $\pm$ corresponds to under- and over- coupled ($K \lessgtr 1$), respectively.  On resonance, full power transfer (critical coupling) from the ``good'' loading channel to the resonant PC cavity mode occurs when $K=1$.  

Whereas $K$ determines the amount of power dropped by the resonator, the role of $I$ is more subtle.  In the case of an internal emitter, the collection efficiency ($\eta_{0}$) of emitted photons into the ``good'' loading channel is,
\begin{equation}\label{eq:eta_0}
\eta_{0} = \frac{\gamma^e_{0}}{\gamma^i_{o} + \sum_{i} \gamma^{e}_{i}} = \frac{1}{1+1/K}.
\end{equation}
\noindent The quality factor of the loaded resonant cavity mode can be written in terms of $K$, $I$, and $Q_{i}$ as $Q_{T}/Q_{i} = 1 - K/(I(1+K)) = 1 - \eta_{0}/I$.  Thus, for a given collection efficiency, to maximize the photon lifetime in the cavity $I$ should be maximized.  For non-linear optical phenomena, where the peak electric-field strength inside the resonant cavity is an important parameter, one can write for the on-resonance internal stored energy
$U = (1 - R_o(\omega_o)) Q_{i+P}  P_{i}$, where $P_{i}$ is the input power in the ``good'' loading channel.  The maximum stored enery in the resonator occurs at $K_{\text{max}} = I/(2-I)$, giving a peak stored energy $U_{\text{max}} = I(Q_{i}/\omega_{o})P_{i}$ which scales directly with $I$.  

The integrated PC cavity-waveguide design employed here has two important features which serve to maximize $I$:  (i) the waveguide and cavity modes have similar transverse field profiles (see Figs.\ \ref{fig:illustration}(b-c)) which allows the cavity to be efficiently loaded end-on, and (ii) the end-fire PCWG-cavity geometry restricts the cavity to a single dominant output channel, in contrast to side-coupled geometries \cite{ref:Noda4} in which the cavity radiates equally into backward and forward propagating PCWG modes (bounding $K < 1$ and $I \le 0.5$).  The geometry in this PC cavity-waveguide system is analogous to a Fabry-Perot cavity with a high reflectivity back mirror and a lower reflectivity front mirror through which a mode-matched input beam sources the cavity.         

\begin{figure}[th]
\begin{center}
\epsfig{figure=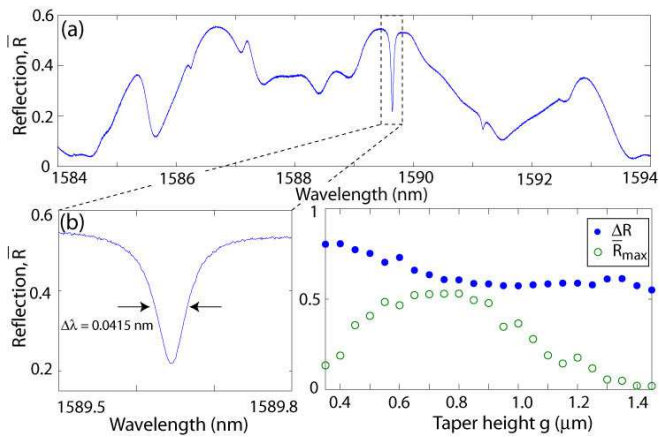, width=1.0\linewidth}
\caption{ (a) Measured reflected taper signal as a function of input wavelength (taper diameter $d \sim 1$ $\mu$m, taper height $g = 0.80$ $\mu$m).  The sharp dip at $\lambda \sim 1589.7$ nm, highlighted in panel (b), corresponds to coupling to the $A_{2}$ cavity mode. (c) Maximum reflected signal (slightly detuned from the $A_{2}$ resonance line), and resonance reflection contrast as a function of taper height.}\label{fig:data}
\end{center}
\end{figure}

In this work, the PC cavity and PCWG devices were fabricated in an optically thin layer (thickness 340 nm) of silicon as described in Ref.\ \cite{ref:Barclay4}.  As shown in Fig.\ \ref{fig:data}, a trench extending diagonally from the cavity was defined to allow direct cavity probing\cite{ref:Srinivasan4}.  Also, for these devices the loading of the PC cavity was set by the 9 periods of air holes between the center of the PC cavity and the end of the PCWG.  A fiber-coupled swept wavelength (1565 - 1625 nm) laser source was used to measure the wavelength dependent forward transmission ($\bar{T}$) through the fiber taper.  The reflected signal in the backward propagating fiber taper mode ($\bar{R}$) was also monitored in order to study the reflective properties of the PC cavity.

With the fiber taper placed above and parallel the PCWG, at phase-matching $\bar{T}$ decreases resonantly as power is coupled from the taper into the PCWG. Coupling to the $TE_{1}$ PCWG mode was verified by studying its dispersive and spatial properties\cite{ref:Barclay4}.  The frequency of the $A_{2}$ cavity mode was also independently determined by probing the cavities directly with the fiber taper as described in Ref.\ \cite{ref:Srinivasan4}.  Two mechanisms were then employed to bring the fiber taper-PCWG coupling bandwidth into resonance with an $A_{2}$ cavity mode.  Coarse tuning was obtained by adjusting, from sample to sample, the nominal hole size and longitudinal lattice constant ($\Lambda_{z}$) of the PCWG.  Fine tuning of the coupler's center wavelength over a 100 nm wavelength range was obtained by adjusting the position, and hence diameter, of the fiber taper region coupled to the PCWG.

Figure \ref{fig:data}(a) shows the normalized reflected fiber signal, $\bar{R}$, for a taper diameter $d\sim1$ $\mu$m which aligns the taper-PCWG coupler bandwidth with that of the $A_{2}$ PC cavity mode.  This signal is normalized to the taper transmission in absence of the PCWG, and since light passes through the taper-PCWG coupler twice, is given by $\bar{R} = \eta_{wg}^2 R_{o}$ (note that both $R_{o}$ and $\eta_{wg}$ are frequency dependent).  In Figure \ref{fig:data}(a), the peak in $\bar{R}$ around $\lambda \sim 1590$ nm corresponds to the phase-matched point of the fiber taper and the $TE_{1}$ PCWG mode.  From the peak value of $\bar{R}_{\text{max}} = 0.53$, a peak taper-PCWG coupling efficiency of $\eta_{wg} \sim 73 \%$ is estimated (where the off-resonant $R_{o}$ is assumed to be unity).  This value is lower than the $97\%$ obtained in previous work\cite{ref:Barclay5} due to coupling to additional higher-order (normal to the Si slab) PCWG modes which interfere with the coupling to the fundamental $TE_{1}$ PCWG mode for strong taper-PCWG coupling.  This can be avoided in future devices by increasing the nominal PCWG hole size relative to that in the PC-cavity or reducing the Si slab thickness, effectively freezing out the higher-order PCWG modes\cite{ref:Barclay4}.

The sharp dip in reflection at $\lambda \sim 1589.7$ nm, shown in detail in Fig.\ \ref{fig:data}(b), corresponds to resonant excitation of the $A_{2}$ PC cavity mode, as confirmed by direct fiber probing of the cavity. The other broad features in $\bar{R}$ correspond to weak Fabry-Perot effects of the PCWG.  The reflected fiber taper signal as a function of taper-PCWG hap height, $g$, is shown in Fig.\ \ref{fig:data}(c). For $g \ge 0.8$ $\mu\text{m}$, $\bar{R}_{\text{max}}$ increases with decreasing $g$ as the coupling from the fiber taper to the $TE_{1}$ PCWG mode becomes stronger.  The reflection contrast, $\Delta R = 1-R_{o}(\omega_o) = (\bar{R}_{\text{max}} - \bar{R}(\omega_o))/\bar{R}_{\text{max}}$, however, remains constant since the PCWG-cavity interaction is independent of the fiber taper to PCWG coupling.  For smaller taper-PCWG gap heights, $g < 0.8$ $\mu$m, fiber taper coupling into higher order PCWG modes and radiation modes becomes appreciable, and  $\bar{R}_{\text{max}}$ decreases for decreased taper height.  The corresponding increase in $\Delta R$ seen in Fig.\ \ref{fig:data}(c) is a result of interference between the $TE_{1}$ mode and higher-order PCWG modes which are excited and collected by the taper, and is not a manifestation of improved coupling between the $TE_{1}$ PCWG mode and the $A_{2}$ PC cavity mode.

From a Lorentzian fit to the $A_{2}$ cavity resonance dip, the normalized on-resonance reflected power is estimated to be $R_o(\omega_{o}) = 0.40$, corresponding to an undercoupled $K=0.225$, with a loaded quality factor of $Q_{T} = 3.8 \times 10^4$.  Substituting these values into Eq.\ (\ref{eq:Q_i+P}) gives for the cavity mode quality factor due to parasitic loading and intrinsic losses, $Q_{i+P} = 4.7 \times 10^4$.  Previous measurements of similar PC cavity devices without an external PCWG load yielded intrinsic quality factors of $4 \times 10^4$ \cite{ref:Srinivasan4}, strongly indicating that the parasitic loading of the PC cavity by the PCWG is minimal, and $I \sim 1$ for this PC cavity-waveguide system.  The input power dependence of the lineshape of the $A_{2}$ PC cavity mode was also investigated, with thermal bistability observed for input powers as small as $1$ mW.  Further details of these measurements will be reported elsewhere. 

The efficiency of power transfer from the fiber taper into the PC cavity is given by, $\eta_{in} = \eta_{wg}\Delta R \approx 44 \%$, which corresponds to the \emph{total} percentage of photons input to the fiber taper which are dropped by the PC-cavity (the fiber taper itself typically has a loss of less than $10\%$).  In the case of an internal emitter light source, the efficiency of collection into the fiber taper for this PC cavity system would be, $\eta_{out} = \eta_{wg}\eta_{0} \approx 13\%$ ($\eta_{0} \approx 18\%$).  Previous measurements of near-ideal coupling between the fiber taper and PCWG\cite{ref:Barclay5} indicate that by adjusting the PCWG as described above, $\eta_{in}$ and $\eta_{out}$ can be increased to $58\%$ and $18\%$, respectively.  More substantially, adjustements in the coupling parameter $K$ towards over-coupling by decreasing the number of air-hole periods between the PC cavity and the PCWG can result in significant increases in $\eta_{in}$ and $\eta_{out}$ with minimal penalty in loaded $Q$-factor for $I \sim 1$.

In conclusion, we have demonstrated an optical fiber coupling scheme to efficiently source and collect light from high-Q ultra-small mode volume PC cavities.  This technique should be useful for future experiments in non-linear optics and cQED, where efficient optical interfacing to planar PC cavities is important.


\begin{thebibliography}{13}
\expandafter\ifx\csname natexlab\endcsname\relax\def\natexlab#1{#1}\fi
\expandafter\ifx\csname bibnamefont\endcsname\relax
  \def\bibnamefont#1{#1}\fi
\expandafter\ifx\csname bibfnamefont\endcsname\relax
  \def\bibfnamefont#1{#1}\fi
\expandafter\ifx\csname citenamefont\endcsname\relax
  \def\citenamefont#1{#1}\fi
\expandafter\ifx\csname url\endcsname\relax
  \def\url#1{\texttt{#1}}\fi
\expandafter\ifx\csname urlprefix\endcsname\relax\def\urlprefix{URL }\fi
\providecommand{\bibinfo}[2]{#2}
\providecommand{\eprint}[2][]{\url{#2}}

\bibitem[{\citenamefont{Srinivasan
  et~al.}(2003{\natexlab{a}})\citenamefont{Srinivasan, Barclay, Borselli, and
  Painter}}]{ref:Srinivasan4}
\bibinfo{author}{\bibfnamefont{K.}~\bibnamefont{Srinivasan}},
  \bibinfo{author}{\bibfnamefont{P.~E.} \bibnamefont{Barclay}},
  \bibinfo{author}{\bibfnamefont{M.}~\bibnamefont{Borselli}}, \bibnamefont{and}
  \bibinfo{author}{\bibfnamefont{O.}~\bibnamefont{Painter}},
  \bibinfo{journal}{Submitted to PRL, available online at
  http://arXiv.org/quant-ph/0309190}  (\bibinfo{year}{2003}{\natexlab{a}}).

\bibitem[{\citenamefont{Lev et~al.}(2004)\citenamefont{Lev, Srinivasan,
  Barclay, Painter, and Mabuchi}}]{ref:Lev}
\bibinfo{author}{\bibfnamefont{B.}~\bibnamefont{Lev}},
  \bibinfo{author}{\bibfnamefont{K.}~\bibnamefont{Srinivasan}},
  \bibinfo{author}{\bibfnamefont{P.~E.} \bibnamefont{Barclay}},
  \bibinfo{author}{\bibfnamefont{O.}~\bibnamefont{Painter}}, \bibnamefont{and}
  \bibinfo{author}{\bibfnamefont{H.}~\bibnamefont{Mabuchi}},
  \bibinfo{journal}{http://arxiv.org/abs/quant-ph/0402093, to appear in
  Nanotechnology}  (\bibinfo{year}{2004}).

\bibitem[{\citenamefont{Srinivasan
  et~al.}(2003{\natexlab{b}})\citenamefont{Srinivasan, Barclay, Painter, Chen,
  Cho, and Gmachl}}]{ref:Srinivasan3}
\bibinfo{author}{\bibfnamefont{K.}~\bibnamefont{Srinivasan}},
  \bibinfo{author}{\bibfnamefont{P.~E.} \bibnamefont{Barclay}},
  \bibinfo{author}{\bibfnamefont{O.}~\bibnamefont{Painter}},
  \bibinfo{author}{\bibfnamefont{J.}~\bibnamefont{Chen}},
  \bibinfo{author}{\bibfnamefont{A.~X.} \bibnamefont{Cho}}, \bibnamefont{and}
  \bibinfo{author}{\bibfnamefont{C.}~\bibnamefont{Gmachl}},
  \bibinfo{journal}{Appl. Phys. Lett.}
  \textbf{\bibinfo{volume}{83}}(\bibinfo{number}{10}), \bibinfo{pages}{1915}
  (\bibinfo{year}{2003}{\natexlab{b}}).

\bibitem[{\citenamefont{Akahane et~al.}(2003)\citenamefont{Akahane, Asano,
  Song, and Noda}}]{ref:Noda4}
\bibinfo{author}{\bibfnamefont{Y.}~\bibnamefont{Akahane}},
  \bibinfo{author}{\bibfnamefont{T.}~\bibnamefont{Asano}},
  \bibinfo{author}{\bibfnamefont{B.-S.} \bibnamefont{Song}}, \bibnamefont{and}
  \bibinfo{author}{\bibfnamefont{S.}~\bibnamefont{Noda}},
  \bibinfo{journal}{Nature} \textbf{\bibinfo{volume}{425}},
  \bibinfo{pages}{944} (\bibinfo{year}{2003}).

\bibitem[{\citenamefont{Gerard}(2003)}]{ref:Gerard3}
\bibinfo{author}{\bibfnamefont{J.-M.} \bibnamefont{Gerard}},
  \emph{\bibinfo{title}{{Solid State Cavity-Quantum Electrodynamics with
  Self-Assembled Quantum Dots}}} (\bibinfo{publisher}{Springer-Verlag},
  \bibinfo{address}{Germany}, \bibinfo{year}{2003}), pp.
  \bibinfo{pages}{269--314}.

\bibitem[{\citenamefont{Brassard et~al.}(2000)\citenamefont{Brassard,
  Lutkenhaus, Mor, and Sanders}}]{ref:Brassard1}
\bibinfo{author}{\bibfnamefont{G.}~\bibnamefont{Brassard}},
  \bibinfo{author}{\bibfnamefont{N.}~\bibnamefont{Lutkenhaus}},
  \bibinfo{author}{\bibfnamefont{T.}~\bibnamefont{Mor}}, \bibnamefont{and}
  \bibinfo{author}{\bibfnamefont{B.}~\bibnamefont{Sanders}},
  \bibinfo{journal}{Phys. Rev. Lett.}
  \textbf{\bibinfo{volume}{85}}(\bibinfo{number}{6}), \bibinfo{pages}{1330}
  (\bibinfo{year}{2000}).

\bibitem[{\citenamefont{Pelton et~al.}(2002)\citenamefont{Pelton, Santori,
  Vuckovic, Zhang, Solomon, Plant, and Yamamoto}}]{ref:Pelton}
\bibinfo{author}{\bibfnamefont{M.}~\bibnamefont{Pelton}},
  \bibinfo{author}{\bibfnamefont{C.}~\bibnamefont{Santori}},
  \bibinfo{author}{\bibfnamefont{J.}~\bibnamefont{Vuckovic}},
  \bibinfo{author}{\bibfnamefont{B.}~\bibnamefont{Zhang}},
  \bibinfo{author}{\bibfnamefont{G.}~\bibnamefont{Solomon}},
  \bibinfo{author}{\bibfnamefont{J.}~\bibnamefont{Plant}}, \bibnamefont{and}
  \bibinfo{author}{\bibfnamefont{Y.}~\bibnamefont{Yamamoto}},
  \bibinfo{journal}{Phys. Rev. Lett.} \textbf{\bibinfo{volume}{89}},
  \bibinfo{pages}{299602} (\bibinfo{year}{2002}).

\bibitem[{\citenamefont{McKeever et~al.}(2004)\citenamefont{McKeever, Boca,
  Boozer, Miller, Buck, Kuzmich, and Kimble}}]{ref:McKeever1}
\bibinfo{author}{\bibfnamefont{J.}~\bibnamefont{McKeever}},
  \bibinfo{author}{\bibfnamefont{A.}~\bibnamefont{Boca}},
  \bibinfo{author}{\bibfnamefont{A.~D.} \bibnamefont{Boozer}},
  \bibinfo{author}{\bibfnamefont{R.}~\bibnamefont{Miller}},
  \bibinfo{author}{\bibfnamefont{J.~R.} \bibnamefont{Buck}},
  \bibinfo{author}{\bibfnamefont{A.}~\bibnamefont{Kuzmich}}, \bibnamefont{and}
  \bibinfo{author}{\bibfnamefont{H.~J.} \bibnamefont{Kimble}},
  \bibinfo{journal}{Science}
  \textbf{\bibinfo{volume}{303}}(\bibinfo{number}{5666}), \bibinfo{pages}{1992}
  (\bibinfo{year}{2004}).

\bibitem[{\citenamefont{Knight et~al.}(1997)\citenamefont{Knight, Cheung,
  Jacques, and Birks}}]{ref:Knight}
\bibinfo{author}{\bibfnamefont{J.}~\bibnamefont{Knight}},
  \bibinfo{author}{\bibfnamefont{G.}~\bibnamefont{Cheung}},
  \bibinfo{author}{\bibfnamefont{F.}~\bibnamefont{Jacques}}, \bibnamefont{and}
  \bibinfo{author}{\bibfnamefont{T.}~\bibnamefont{Birks}},
  \bibinfo{journal}{Opt. Lett.}
  \textbf{\bibinfo{volume}{22}}(\bibinfo{number}{15}), \bibinfo{pages}{1129}
  (\bibinfo{year}{1997}).

\bibitem[{\citenamefont{Barclay
  et~al.}(2003{\natexlab{a}})\citenamefont{Barclay, Srinivasan, and
  Painter}}]{ref:Barclay2}
\bibinfo{author}{\bibfnamefont{P.~E.} \bibnamefont{Barclay}},
  \bibinfo{author}{\bibfnamefont{K.}~\bibnamefont{Srinivasan}},
  \bibnamefont{and} \bibinfo{author}{\bibfnamefont{O.}~\bibnamefont{Painter}},
  \bibinfo{journal}{J. Opt. Soc. Am. B}
  \textbf{\bibinfo{volume}{20}}(\bibinfo{number}{11}), \bibinfo{pages}{2274}
  (\bibinfo{year}{2003}{\natexlab{a}}).

\bibitem[{\citenamefont{Barclay et~al.}(2004)\citenamefont{Barclay, Srinivasan,
  Borselli, and Painter}}]{ref:Barclay5}
\bibinfo{author}{\bibfnamefont{P.~E.} \bibnamefont{Barclay}},
  \bibinfo{author}{\bibfnamefont{K.}~\bibnamefont{Srinivasan}},
  \bibinfo{author}{\bibfnamefont{M.}~\bibnamefont{Borselli}}, \bibnamefont{and}
  \bibinfo{author}{\bibfnamefont{O.}~\bibnamefont{Painter}},
  \bibinfo{journal}{Opt. Lett.}
  \textbf{\bibinfo{volume}{29}}(\bibinfo{number}{7}), \bibinfo{pages}{697}
  (\bibinfo{year}{2004}).

\bibitem[{\citenamefont{Spillane et~al.}(2003)\citenamefont{Spillane,
  Kippenberg, Painter, and Vahala}}]{ref:Spillane2}
\bibinfo{author}{\bibfnamefont{S.~M.} \bibnamefont{Spillane}},
  \bibinfo{author}{\bibfnamefont{T.~J.} \bibnamefont{Kippenberg}},
  \bibinfo{author}{\bibfnamefont{O.~J.} \bibnamefont{Painter}},
  \bibnamefont{and} \bibinfo{author}{\bibfnamefont{K.~J.}
  \bibnamefont{Vahala}}, \bibinfo{journal}{Phys. Rev. Lett.}
  \textbf{\bibinfo{volume}{91}}(\bibinfo{number}{4}), \bibinfo{pages}{043902}
  (\bibinfo{year}{2003}).

\bibitem[{\citenamefont{Barclay
  et~al.}(2003{\natexlab{b}})\citenamefont{Barclay, Srinivasan, Borselli, and
  Painter}}]{ref:Barclay4}
\bibinfo{author}{\bibfnamefont{P.~E.} \bibnamefont{Barclay}},
  \bibinfo{author}{\bibfnamefont{K.}~\bibnamefont{Srinivasan}},
  \bibinfo{author}{\bibfnamefont{M.}~\bibnamefont{Borselli}}, \bibnamefont{and}
  \bibinfo{author}{\bibfnamefont{O.}~\bibnamefont{Painter}},
  \bibinfo{journal}{To appear in Appl. Phys. Lett., available online at
  http://arXiv.org/quant-ph/0308070}  (\bibinfo{year}{2003}{\natexlab{b}}).

\end{thebibliography}
\end{document}